\documentclass[english,keywords,amsmath,amssymb,twocolumn]{revtex4}
\usepackage[T1]{fontenc}
\usepackage[latin1]{inputenc}
\usepackage{babel}%
\usepackage{graphicx}
\usepackage{color}
\usepackage{bm}
\usepackage{longtable}
\usepackage{amsmath}
\usepackage{amsfonts}
\usepackage{amssymb}

\begin{document}
\title{Comment on ``Weak Measurements with Orbital-Angular-Momentum Pointer states''}
\author{A. K. Pan$^{1}$}
\author{P. K. Panigrahi$^{2}$}
\affiliation{$^{1}$Graduate School of Information Science, Nagoya University, Chikusa-ku, Nagoya 464-8601, Japan}
\affiliation{$^{2}$Indian Institute of Science Education and Research Kolkata, Mohanpur, Nadia 741252, India}
\maketitle
In their recent Letter\cite{puentes}, Puentes \emph{et al.} have provided a scheme for extracting the real and imaginary parts of higher-order weak value by using orbital angular momentum(OAM) states as pointer states, based on two interaction Hamiltonians and multiple joint pointer displacements. They claim that such weak values are inaccessible with Gaussian pointer state only, due to its particular symmetry property. As explained in \cite{puentes}, OAM states introduce a `different symmetry' for the expectation values of operators, and thus provide access to such weak values. The purpose of this Comment is to show that for the same Hamiltonian, Gaussian pointer state by itself can provide access to the real part of higher-order weak value, if suitable pointer displacement is observed. This is because the symmetry property responsible for revealing such value can also be restored with Gaussian pointer state for a suitable choice of the measurement of the observable after post-selection.  

Puentes \emph{et al.}\cite{puentes} considered a joint weak measurement of two commuting observables $\hat A$ and $\hat B$ satisfying  $(\hat A)^2\neq I$ and $ (\hat B)^2 \neq I$. The system-apparatus state at $t=0$ is $ |\Psi_{i}\rangle=|\psi_{i}\rangle|i\rangle=\int dx dy  \psi(x,y)|x,y\rangle  |i\rangle$ where $|i\rangle$ denotes pre-selected system state, and the pointer wave function $\psi(x,y)$ is taken to be Laguerre-Gauss mode 
\begin{eqnarray}
\label{lg}
\psi(x,y)=N(x+i sgn(l) y)^{|l|}exp\left(-\frac{x^2 +y^2}{4\sigma^2}\right),
\end{eqnarray}
where $N= 1/\sqrt {\pi 2^{l+1}\sigma^{2l+2} l!}$ being the normalization constant. For $l=0$, Eq.(\ref{lg}) reduces to Gaussian one.

The interaction Hamiltonian is taken to be $\mathcal{H}_{I}=g(t)(\hat A  \hat P_x + \hat B \hat P_y)$, where $\hat P_{x}$ and $\hat P_y$ are the conjugate momenta of the position observables $\hat X$ and $\hat Y$ respectively. The quantity $g(t)$ obeys $\int_{0}^{t}g(t)dt= g$, and $t$ is the duration of interaction. The time evolved state after interaction between system and probe is,$|\psi_{f}\rangle=\int dx dy \  e^{-i g(\hat A \hat P_x + \hat B \hat P_y)}|i\rangle\psi(x,y)|x,y\rangle$.

Similar to Ref.\cite{puentes}, expanding the exponential up to the second order of $g$, and post-selecting the system in a state $|f\rangle$, the final pointer state,
\begin{eqnarray}
\label{puen1}
|\psi_{f}\rangle=\int dx dy \mathcal{Z} \psi(x,y)|x,y\rangle  |\langle f|i\rangle
\end{eqnarray}
where $
\mathcal{Z}=1-i g (\hat A)_w \hat  P_x -i g  (\hat B)_w \hat P_y 
-g^2  (\hat A^2)_w \hat  P_x^2/2 - g^2  (\hat B^2)_w \hat P_y^2/2 -g^2  (\hat A \hat B)_w \hat  P_x \hat P_y
$ with $(..)_{w}=\langle f|..|i\rangle/\langle f|i\rangle$ denoting the standard definition of weak value\cite{ah}.

Using Eq.(\ref{puen1}), Puentes \emph{et al.}\cite{puentes} calculated the pointer displacement of the joint meter observable $\hat X\hat Y$,
\begin{eqnarray}
\label{puen2}
\langle \hat X \hat Y\rangle_{f}&=&\langle \hat X \hat Y\rangle_{i}+\frac{g^2 }{2}  \left(\Re[(AB)_{w}]+\Re[(A)_{w}^{\ast}B_{w}]\right)\\
\nonumber
&+&\frac{g^2}{2}\left[l\left(\Im[A^2_{w}]+\Im[B^2_{w}]\right)\right]
\end{eqnarray}
implying the imaginary part of the weak values of higher-order moments of $\hat A$ and $\hat B$ are dependent on $l$. The quantities $\langle \hat X\hat P_{y}\rangle$ and $\langle \hat Y\hat P_{x}\rangle$ are also calculated for obtaining real parts. This is the main result of Ref.\cite{puentes}. 

Now, for $l=0$, the higher-order weak values in Eq.(\ref{puen2})disappear leading them to conclude that such values cannot be accessible with Gaussian pointer state. 

Here we show that such weak values are also accessible with Gaussian pointer state for the same Hamiltonian.  For this, instead of joint pointer displacement used in \cite{puentes}, we consider the displacement of single pointer observable $\hat X^2$. By putting $l=0$ in Eq.(\ref{lg}), and using Eq.(\ref{puen1}), the expectation value of $\hat X^2$, is calculated as, 
\begin{eqnarray}
\label{final}
&&\langle \hat X^2\rangle_{f}=\\
\nonumber
&&\left[\langle \hat X^{2} \rangle_{i} +\frac{g^2 }{2} (|B_w|^2-|A_w|^2 +\Re[A^2_{w}]-\Re[B^2_{w}])\right] W^{-1}
\end{eqnarray}
where $W=\langle\psi_{f}|\psi_{f}\rangle/(|\langle f|i\rangle|^2)$, and $\langle \hat X^{2} \rangle_{i}$ is constant quantity, depends only on initial width of pointer state.

It is evident from Eq.(\ref{final}) that real part of higher-order weak values can be obtained with Gaussian pointer state for the same Hamiltonian used in \cite{puentes}.  However, to access the imaginary part with Gaussian pointer state, a different interaction Hamiltonian $\mathcal{H}_{I}= g_{A}(t)A\otimes p_{x}+ g_{B}(t) B\otimes x$(used in \cite{wu}) and the displacement of  pointer observable $(\hat X \hat P_{x} +\hat P_{x} \hat X)/2$, need to be considered. Furthermore, while $B\equiv 0$ in the first Hamiltonian, higher-order weak value $\Re[(A^2)_w]$, and related signal amplification can be achieved by using one-dimensional Gaussian pointer state only, thereby making the scheme simpler. We thus demonstrated that the Gaussian pointer state by itself can provide access to the higher-order weak value as opposed to the claim in \cite{puentes}. \\

Acknowledgments: AKP acknowledges the support from JSPS Postdoctoral Fellowship for Foreign Researcher and Grant-in-Aid for JSPS fellows no. 24-02320.

\end{document}